# $UHg_3$ – A Heavy Fermion Antiferromagnet Similar to $U_2Zn_{17}$ and $UCd_{11}$

J. S. Kim and G. R. Stewart[*]

Department of Physics, University of Florida, Gainesville, FL 32611.

**Abstract**: Heavy Fermion physics deals with the ground state formation and interactions in f-electron materials where the electron effective masses are extremely large, more than 100 times the rest mass of an electron. The details of how the f-electrons correlate at low temperature to become so massive lacks a coherent theory, partially because so few materials display this 'heavy' behavior and thus global trends remain unclear. $UHg_3$ is now found experimentally to be a heavy Fermion antiferromagnet – just as are *all* the other $U_xM_y$ compounds with the metal M being in column IIB (filled d-electron shells) in the periodic table (Zn/Cd/Hg) and the spacing between Uranium ions, $d_{U-U}$, being greater than the Hill limit of 3.5 Å. This result, that - independent of the structure of these $U_xM_y$, M=Zn/Cd/Hg, compounds and independent of the value of their $d_{U-U}$ (ranging from 4.39 to 6.56 Å) – all exhibit heavy Fermion antiferromagnetism, is a clear narrowing of the parameters important for understanding the formation of this ground state. The sequence of antiferromagnetic transition temperatures, $T_N$, of 9.7 K, 5.0 K, and 2.6 K for $U_xM_y$ as the metal M varies down column IIB (Zn/Cd/Hg) indicates an interesting regularity for the antiferromagnetic coupling strength.

**Main Text:**

The study of 'heavy Fermion' physics became[1] a recognized sub-field of condensed matter physics starting with the discovery in 1979 of superconductivity at $T_c = 0.6$ K in $CeCu_2Si_2$, a material with a large ('heavy') effective mass, m*, of the paired electrons – greater than 100 times that of a regular superconducting metal like Sn or Al. By 1984, two additional superconductors, two non-ordering compounds, and three antiferromagnets had been added to the list of heavy Fermion systems.

The present work focuses on a discovery that may help to explain the sub-class represented by the two antiferromagnets, $U_2Zn_{17}$ and $UCd_{11}$, whose correlated density of states at the Fermi energy (proportional to the linear term γ coefficient in the specific heat C/T (T→0)=γ+ßT², where m* ∝ γ)) is *independent* of temperature at low temperature. This is in contrast with the other six compounds, which show specific heat γ's rising between 50 and 500% with decreasing temperature below 10 K. Thus, these six strongly correlated electron systems do not re-enter the renormalized Fermi liquid state (where γ would be constant) at even very low temperature, i. e. they remain 'non-Fermi liquid' like in the sense that the strong correlations responsible for the γ values are still temperature dependent as T→0.

Using phenomenological arguments involving the Hill limit[2] and the known behavior of $U_2Zn_{17}$ and $UCd_{11}$ as the basis for prediction, we report the third large (defined in ref. 1 as γ>400 mJ/molK$^2$) <u>constant-γ</u> heavy Fermion compound (again an antiferromagnet), $UHg_3$. This prediction/experimental work is an example of successful 'materials by design', and provides focused input into the theory of understanding heavy Fermion ground state formation – at least in the case where γ is a constant at low temperature.

The choice of $UHg_3$ for investigation of possible heavy Fermion antiferromagnetic behavior involved the following considerations. The first two elements in column IIB of the periodic table, Zn and Cd, form[1] binary heavy Fermion antiferromagnetic compounds with U ($U_2Zn_{17}$, $T_N$=9.7 K, with the spacing between the U atoms, $d_{U-U}$, = 4.39 Å and $UCd_{11}$, $T_N$=5.0 K, $d_{U-U}$=6.56 Å) (γ=535 and 840 mJ/UmolK$^2$ respectively). These γ values imply effective electron masses greater than 100 that of a normal metal and are proportional to the correlated density of states at the Fermi energy, $N(\varepsilon_F)*(1 + \lambda)$. Since γ at low temperatures in $U_2Zn_{17}$ and $UCd_{11}$ is constant with temperature at low temperatures, this implies that the mass renormalization due to the (1+λ) factor is also temperature independent at low temperatures, i. e. the samples have entered into a new, renormalized Fermi liquid ground state upon cooling. Such a simplifying assumption – that the mass renormalization is constant at low temperatures in these two systems

– is an important input for any theory explaining this highly correlated ground state. Thus, if a third such system could be found, this could serve to provide useful input to theory for understanding highly correlated metals physics.

In order to put the rarity of materials known to have such large γ values in perspective, prior to the present work $U_2Zn_{17}$ and $UCd_{11}$ were the only two known systems with <u>constant-γ</u> values > 400 mJ/molK$^2$ at low temperatures. If we consider all known systems with 2, 3 or 4 constituent elements and including superconductors, magnets, and non-ordered systems like $CeCu_6$ and $CeAl_3$, there are less[1,3] than 15 known with temperature *dependent* γ values (i. e. where the γ is due to temperature dependent spin fluctuations, perhaps from a quantum critical point[4-5]) above 400 mJ/molK$^2$. Since the definition of how large a γ has to be to be named "heavy" Fermion is arbitrary (although note that simple d-electron materials such as ß-manganese have[6] γ values of 60 mJ/molK$^2$), if the range is extended down to 200 mJ/molK$^2$, there are still only approximately 30 systems, including the antiferromagnets $NpIr_2$ (γ=230 mJ/molK$^2$)[7], $UZn_{12}$ (γ=200 mJ/molK$^2$)[8] and $UMn_2Al_{20}$ (γ=200 mJ/molK$^2$)[9]. The normal state (T>$T_N$) γ values in these three materials are, although a factor of 2.5-4 smaller than in $U_2Zn_{17}$ and $UCd_{11}$, also temperature independent at low temperatures.

It is important to note here that there *must* be hybridization taking place between the U 5 f electrons and the ligand electrons in the column IIB atoms Zn and Cd in the heavy Fermion antiferromagnets $U_2Zn_{17}$ and $UCd_{11}$. This U 5f – Zn/Cd ligand electron hybridization has to be present since the distance between neighboring U atoms in both cases is larger than the Hill limit[2] of 3.5 Å. Below the Hill limit, the 5 f electron orbitals can overlap and form conduction bands like in a regular metal. Beyond the 3.5 Å limit the 5 f orbitals are either localized (where localized electrons are not available to contribute to the enhanced density of states at the Fermi energy, $N(\varepsilon_F)*(1 + \lambda))$ *or* are at least partially itinerant due to hybridization with intervening ligand atom electrons.  There are no heavy Fermion compounds where the f electron orbitals overlap – the mechanism for forming a highly correlated, high effective mass ground state depends on the f electron-ligand electron hybridization.  The theory of how this hybridization functions in detail needs further development, since $U_2Zn_{17}$ and $UCd_{11}$ have differing structures and symmetries (rhombohedral and cubic respectively) and differing $d_{U-U}$, but yet rather similar (in the sense of similar size $\gamma$ and $T_N$ values and both with temperature independent $\gamma$ values) heavy Fermion antiferromagnetism.

The present work investigated whether this quasi-unique hybridization between U 5 f electrons and the column IIB ligand Zn and Cd electrons would translate into a heavy Fermion antiferromagnet if a $U_xHg_y$ compound could be found with $d_{U-U}>3.5$ Å, where Hg is the third element after Zn and Cd in column IIB.  In support of this premise, the only other compound present in the U-Zn phase diagram besides the heavy Fermion antiferromagnet $U_2Zn_{17}$ is $UZn_{12}$ (an antiferromagnet with $d_{U-U}\geq 4.45$ Å and an intermediate size $\gamma$ of 200 mJ/molK$^2$)[8] and there is only the $UCd_{11}$ compound in the U-Cd phase diagram.  Thus, based on these two members of the possible column IIB ligands with U, the idea that $U_xHg_y$ will be a heavy Fermion antiferromagnet if a $U_xHg_y$ compound with $d_{U-U}>3.5$ Å can be found served as the phenomenological basis for the present work.

In the U-Hg phase diagram, preparation and characterization of the samples is difficult since[10] all the compounds "oxidize with great rapidity."  This difficulty explains the dearth of characterization of $U_xH_y$ compounds: magnetic susceptibility, $\chi$, data on[11] $UHg_2$ exist and resistivity data have been measured on $UHg_3$ to infer[12] "some kind of ordering" near 50 K. The known $U_xHg_y$ compounds have the following structures and $d_{U-U}$ values. $UHg_2$ is a regular metal with overlapping f-orbitals, $d_{U-U}$=3.22 Å, antiferromagnetic at[11] ~ 70 K with a hexagonal unit cell with 3 atoms. $UHg_3$ has been determined[12] to have a hexagonal structure which is however "unrefined", with therefore an unknown $d_{U-U}$. Ref. 10 discusses the placement of the U atoms in $UHg_3$ as randomly distributed over the position of hexagonal closest packing. There are a number of $MHg_3$ compounds (with, e. g., M=Ce, La, Lu, Sc, Y, Yb) which occur in the hexagonal structure (called 'DO19') in which two well known heavy Fermion compounds ($UPt_3$ and $CeAl_3$) also form, with 8 atoms per unit cell. If the $d_{U-U}$ in $UHg_3$ were to be calculated using the placement of atoms in this DO19 hexagonal structure, then $d_{U-U}$~4.54 Å. In any case, if heavy Fermion behavior is found in $UHg_3$, this would be a proof that $d_{U-U}$>3.5 Å, the Hill limit. The third compound in the binary U-Hg phase diagram is $U_{11}Hg_{45}$, which has $d_{U-U}$=5.3 Å and a cubic face centered unit cell with 448 atoms.

Thus, below, we discuss the preparation and characterization of $UHg_3$. Just as in the U-Zn phase diagram where both $U_2Zn_{17}$ and $UZn_{12}$ have large γ values (535 and 200 mJ/UmolK$^2$ respectively), $U_{11}Hg_{45}$ may also be of interest. However, in the constrained vapor (P~100 atm) phase diagram $UHg_3$ is stable up to 735 °C, where it forms in a peritectic reaction, while $U_{11}Hg_{45}$, which also forms peritectically, is only stable up to 455 °C. Thus, $UHg_3$ should form more readily and characterization thereof will provide a test of our prediction.

Samples of $UHg_3$ were prepared[13] using 60 mesh U powder, 99.7 % pure, which has been deoxidized using 1:1 HNO3 and water. The Hg used was 99.999995% pure. Stoichiometric amounts of the U and Hg were placed in a pre-outgassed alumina crucible with lid, which was sealed in a Nb cylinder with a bottom and top Nb lid welded on. This was then placed in a tube furnace through which Ar flowed to prevent the Nb from oxidizing, heated to

900 °C, held for 10 hours, and cooled at 5 °C/hr to 250 °C, followed by 75 °C/hr cooling to room temperature. The amount of Hg in the approximately 6 cm$^3$ volume Nb containment was kept below about 200 mg. Even so, the pressure generated inside was sufficient to bow the flat endcaps outwards slightly during the reaction sequence. There was no reaction between the Hg and the containment. In addition, all of the Hg was reacted with the U powder, with no excess remaining. X-ray diffraction characterization revealed less than 5% of any possible second phase (e. g. UHg$_2$, U$_{11}$Hg$_{45}$, or U metal). The x-ray diffraction peaks of the prepared UHg$_3$ are similar to those of the DO19 hexagonal structure, but with several weak peaks (e. g. [102] and [103]) in the DO19 pattern either extinct or much reduced in intensity. As has been pointed out[10], U and Hg have very similar ionic radii which could lead to site switching. However, the Z values (92 and 80 of U and Hg respectively) and scattering factors are rather similar. Thus, our calculations of the x-ray diffraction pattern using the program Powder Cell indicate that such site switching cannot explain the disagreement between a calculated DO19 x-ray pattern and that measured in the present work for UHg$_3$. Thus, UHg$_3$ appears not to have a disordered DO19 hexagonal structure.

The magnetic susceptibility, $\chi$, of UHg$_3$ is shown in Fig. 1. As will be seen below when the specific heat data are presented, there are two bulk anomalies at 2.6 and 4.8 K respectively. Magnetization vs field up to 5 T (not shown) is essentially linear up to 5 T for the sample of UHg$_3$ presented here, implying no magnetic impurities present to exhibit saturation behavior. This linearity of M vs H, combined with the rather large value of $\chi$ (2 K) (13 memu/mol)), is consistent[14] with heavy Fermion behavior. $\chi$ at low temperatures for U$_2$Zn$_{17}$ is 4.5 memu/Umol and for UCd$_{11}$ is 38 memu/mol.[1]

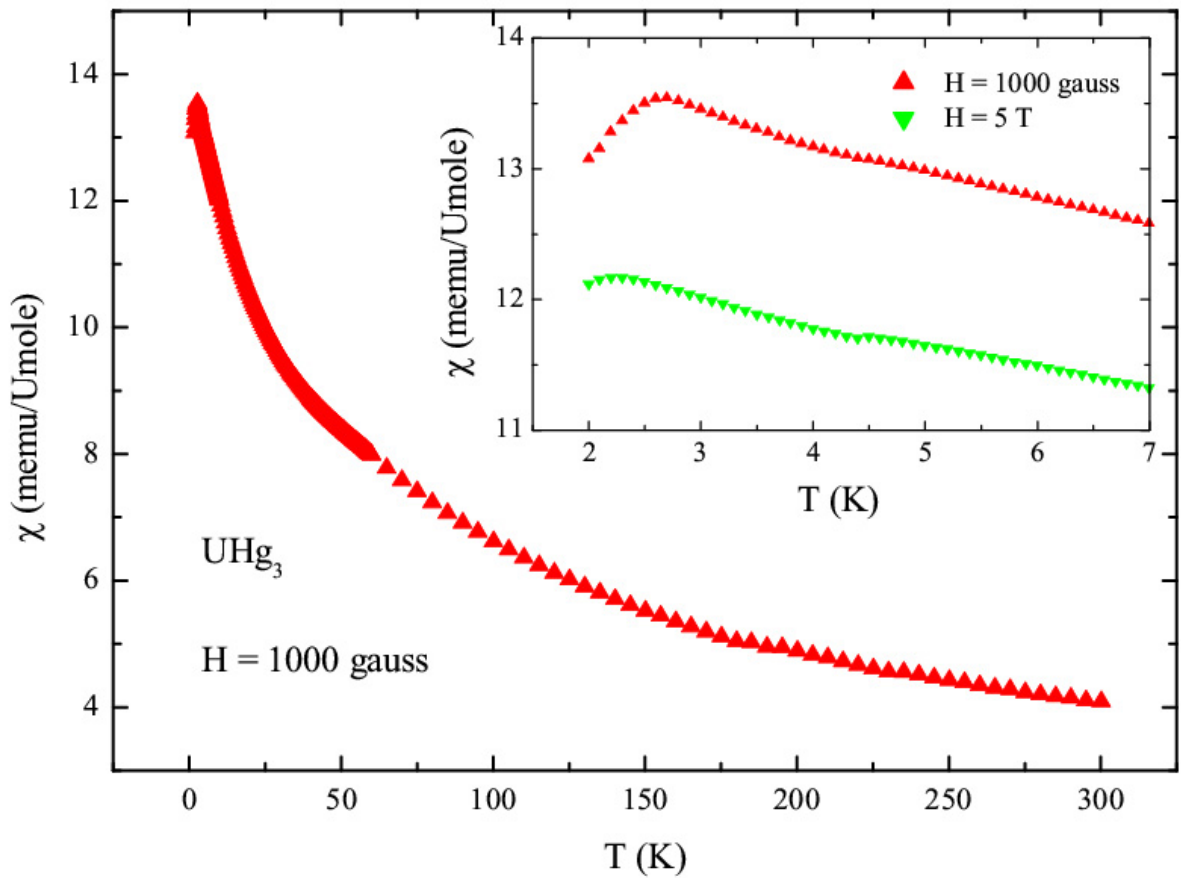

Figure 1 (color online) The dc magnetic susceptibility of polycrystalline UHg$_3$ both on an expanded low temperature scale at two applied fields and over the whole 2-300 K measurement range in 1000 Gauss. Note that the peak in $\chi$ at 2.7 K and its shift downwards with a 5 T field are both consistent with an antiferromagnetic transition. There is also a slight ($\approx$ 2%) anomaly at 4.5 K, see the specific heat in Fig. 2. This slight anomaly shows no field dependence in $\chi$ up to 5 T. The small change in slope of $\chi$ around 45 K may correspond to the anomaly observed in the resistivity in ref. 12. If a constant term (2.2 memu/Umole) is subtracted from these $\chi$ data, a plot (not shown) of 1/$\chi$ vs T is linear and implies an effective moment of 2.3 $\mu_B$ per U atom, consistent with an f$^1$ state for the U ion.

The specific heat of UHg$_3$ at low temperatures down to 0.4 K is presented in Fig. 2. The zero field data are plotted by themselves in Fig 2a to accentuate the good linearity of C/T plotted vs T$^2$, i. e. the extrapolation shown is clearly convincing evidence for a $\gamma$ value that is constant at these low temperatures.. The peak in $\chi$ at 2.7 K shown in Fig. 1 clearly correlates with the large, sharp peak in the specific heat at 2.35 K (onset at 2.8 K) in Fig. 2. Also, the field suppression of the ordering temperature measured by the field specific heat data is further consistent with this anomaly indeed being an antiferromagnetic transition.

The anomaly in the specific heat starting at around 5.3 K, on the other hand, shows no field dependence up to the highest field of measurement, 12 T. This anomaly, as well as the antiferromagnetic transition at $T_N^{mid}$ = 2.6 K, was present in all seven of the UHg$_3$ samples measured, including those made from small chunks (similar to the method in ref. 16 for preparing CeHg$_3$) of U rather than from U powder, and is clearly an intrinsic feature.

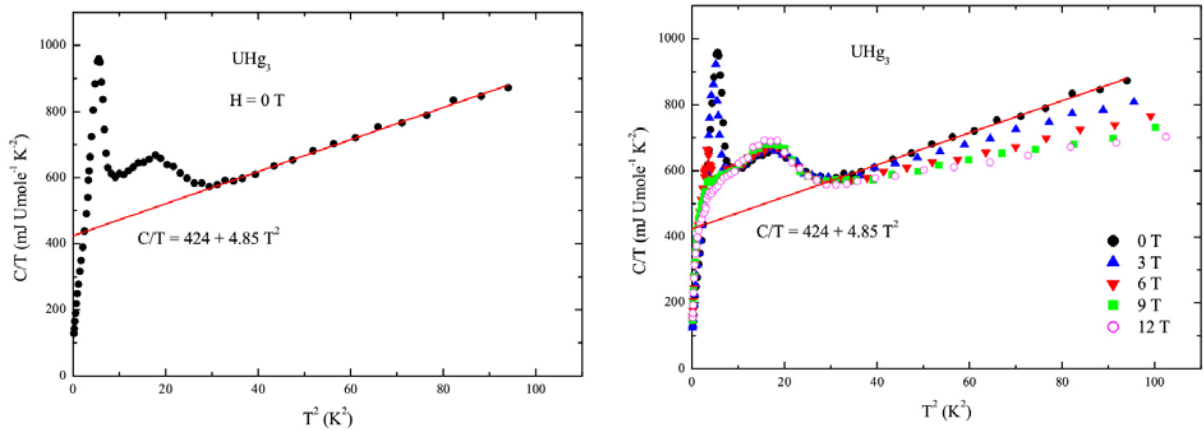

Figure 2a,b (color online) Specific heat, C, divided by temperature, T, vs $T^2$ in zero and applied field up to 12 T between 0.4 and 10 K of pressed pellets of polycrystalline UHg$_3$. The peak in C/T is at 2.35 K, while the onset of this transition is about 2.8 K, giving a midpoint of about 2.6 K. The shoulder in the specific heat starts at around 5.3 K and peaks at 4.3 K. The zero field $T^2$ coefficient for C/T (4.85 mJ/molK$^2$) gives a Debye temperature, $\theta_D$, of 117 K, indicating a rather soft lattice. The extrapolation of the C/T data from above (starting at T=5.9 K) the upper transition – which are quite linear when plotted vs $T^2$ - gives a value for γ of over 400 mJ/molK$^2$. This value is robust, varying by <3 % if a different starting temperature (6.9 vs 5.9 K) is chosen for the fit of C/T to γ + β$T^2$. The low temperature extrapolation of the C/T data in the antiferromagnetic ordered state is only 35 mJ/molK$^2$, a large reduction from the high temperature extrapolated value. Note that the applied magnetic field data in Fig. 2b imply an increase of γ with field, consistent with the results[15] of the specific heat in field of UPt$_3$, where spin fluctuations are present.

Without low temperature x-ray characterization or neutron scattering data the nature of the ~5 K, field-independent transition in the specific heat is presently uncertain. It is worthwhile to note however that the upper transition looks very similar to the structural transition (cubic →tetragonal upon cooling) observed[17-18] in the specific heat in the A-15 structure superconductors $Nb_3Sn$ and $V_3Si$. In $V_3Si$ the depression with field of this structural transition (at 21.3 K, while the superconducting transition is at 16.9 K) was found[17] to be quite small, 0.26 K in 9 T. Also, there is[19] a 6% decrease in χ as $Nb_3Sn$ is cooled through its structural transition at 45 K (vs 2% in the data for $UHg_3$ shown in Fig. 1) due to a change in the electronic density of states at the Fermi energy caused by the change in structure.

Finally, the size of the specific heat γ (where C/T is fit to $\gamma + \beta T^2$) extrapolated from above 5.5 K is, as shown in Fig. 2, approximately 420 mJ/UmoleK$^2$. This value is – together with the presence of antiferromagnetism at 2.6 K – validation of the prediction of heavy Fermion antiferromagnetism in $U_xHg_y$. Although there are certainly differences (e. g. the decrease in γ below the antiferromagnetic transitions in $U_2Zn_{17}$ and $UCd_{11}$ is[1] 63 and 70% respectively vs 92% in $UHg_3$ – i. e. the antiferromagnetic transition's reconstruction of the Fermi surface is much more severe in $UHg_3$), the series of heavy Fermion antiferromagnets $U_2Zn_{17}$ ($T_N$ = 9.7 K), $UCd_{11}$ ($T_N$=5.0 K), and $UHg_3$ ($T_N$=2.6 K) must be considered to belong to a related family of compounds. Certainly, if one looks at a log-log plot of γ vs χ as put forward by Z. Fisk and also by B. Jones (see ref. 20), the points for these three materials cluster rather closely together.

The question for which new theory is now needed is: how does the hybridization of the U 5 f electrons with the electrons in the column IIB elements Zn, Cd, and Hg cause this quasi-unique, <u>constant-γ</u> heavy Fermion antiferromagnetism? The details of this outer shell ligand electron hybridization of the column IIB elements with the U 5 f electrons needs to be addressed theoretically. Also, is the decreasing magnitude of $T_N$ as M in $U_xM_y$ shifts down the column IIB elements (approximately a factor of two between each member, with a second value for M=Zn of

$T_N = 5.0$ K for UZn$_{12}$) accidental, or also understandable through a proper theoretical explanation?

**Acknowledgements:** Work at Florida supported by the U. S. Department of Energy, Office of Basic Energy Sciences, Division of Materials Sciences and Engineering under Award no. DE-FG02-86ER45268. The authors thank Bohdan Andraka for the uranium powder, David Tanner for continuing use of his inert atmosphere glove box, and Pradeep Kumar for useful theoretical discussion.